\documentclass[11pt]{article}
\usepackage{epsfig}
\usepackage{verbatim}
\usepackage[pdfmenubar=false, letterpaper=false, pdfpagelabels=true]{hyperref}
\begin{document}

\title{About the proof of the so called \\
{\it exact classical confidence intervals}.\\
Where is the trick?\,\footnote{Based on a 
lectures to graduate students at the University of Rome ``La Sapienza''
(May, 4 2005 and May, 8 2006).}}
\author{G.~D'Agostini \\
Universit\`a ``La Sapienza'' and INFN, Roma, Italia \\
{\small (giulio.dagostini@roma1.infn.it,
 \url{http://www.roma1.infn.it/~dagos})}
}

\date{}

\maketitle

\begin{abstract}
In this note I go through the `proof' of frequentistic
confidence intervals and show what it logically 
implies concerning the value of a physical quantity
given an experimental observation (nothing).
\end{abstract}
\vspace{1.2cm}
{\small
\begin{flushright}
{\sl ``\ldots to emancipate us from the } \\
{\sl capricious ipse dixit of authority'' } \\
{(John Henry Newman)}\\
\end{flushright}
}

\section{Introduction}

The construction of frequentistic confidence intervals
is the statistical tool adopted
in all `conventional statistics'~\cite{Maxent98} books and lecture notes
in order to provide the result on the value of a physical
quantity resulting from a measurement. 
Though many physicists are aware of the
unsuitability of the `prescription' 
to handle critical cases that occur in
frontier science (and the troubles are also known to 
the supporters of the prescription: see \cite{clw}, 
references therein and follows up), it seems they do not
always recognize
the real reason the prescription fails. 

In the past I am always been reluctant to go 
through the details of the definition of the 
frequentistic confidence intervals, 
simply because I thought that, once one has realized that:
\begin{itemize}
\item
the outcome of these methods is usually misleading
(in the sense that the common interpretation differs from what they are
supposed to mean, see e.g. Ref.~\cite{Maxent98});
\item
the resulting `confidence' intervals can often come out absurd
(see e.g. Ref.~\cite{BR}, sections 1.7 and 10.7);
\item
the celebrated `frequentistic coverage' does not do its job
in the important cases of interest in frontier physics 
(see Ref.~\cite{BR}, section 10.7);
\end{itemize}
then there is little more to do, apart from looking for something
better. I admit I have been naive. 
In fact I have met too many students and colleagues touched
by the above arguments
but still not fully convinced, because  
impressed by the names given to the 
criticized methods (something that is
{\it classical},  {\it rigorous} and {\it exact} cannot be wrong!) 
and the authority of 
books and publications that use and recommend 
them.\footnote{I found a curious analogy between 
the learning of fundamental concepts of physics and the 
learning of statistical methods for data analysis, as they are taught
in the standard physics curriculum. It is a matter of fact
that many concepts of physics are not easy at all to be grasped
by students (and not not only by students!). After some time,
students assume the habit
to learn at least the practical
formulas, trusting teachers and text books
as far as correctness and deep meaning
of the difficult concepts are concerned.
When they first learn frequentistic statistics applied to
data analysis, presented in the usual shifty way we know,
the students tend to act in the same way,
thinking the things must be difficult but correct,
because granted by the teachers,
who in most case just repeat the lesson they learned but tacitly
hold the same doubts of the pupils. The small `Socratic exchange'
by George Gabor reported in Ref.~\cite{BR} might be 
enlightening.
}

The aim of this note is 
to go through the definition of the confidence interval as one
finds it in books and lecture 
notes\footnote{I avoid to give a particular reference.
Everyone can check his/her preferred text and see how
confidence intervals are presented.}
and to show why that reasoning yields no new information 
concerning  the value of the unknown quantity 
of interest and about which one wants to gain some confidence 
based on experimental observations.
\newpage
\section{From the probability interval of the estimator $\hat\theta$ 
to the confidence about the true value $\theta$}
Here is, essentially,
the basic reasoning that introduces confidence
intervals. 
\begin{enumerate}
\item Let $\theta$ be the {\it true value} of the quantity 
      of interest and $\hat\theta$ its `estimate' (an experimental
      quantity, probabilistically related to $\theta$). 
\item We assume to know, somehow,\footnote{I skip over the fact that 
      in most cases $f(\hat\theta\,|\,\theta)$ is not really obtained 
      by past frequencies
      of $\hat\theta$ for some fixed value of $\theta$, as the frequency based
      definition of probability would require.} 
      the probability density function (pdf) of 
      $\hat\theta$ for any given value 
       $\theta$, i.e. $f(\hat\theta\,|\,\theta)$.
       The knowledge of $f(\hat\theta\,|\,\theta)$ allows 
      then us to make probabilistic
      statements about $\hat\theta$, for example that $\hat\theta$ will occur
      in the interval between $\theta-\Delta\theta_1$ and $\theta+\Delta\theta_2$:
      \begin{eqnarray}
      P(  \theta-\Delta\theta_1    < \hat\theta < \theta+\Delta\theta_2) &=& \alpha\,.
      \label{eq:P_est}
      \end{eqnarray}
      For example, if $f(\hat\theta\,|\,\theta)$ is a Gaussian and 
      $\Delta\theta_1=\Delta\theta_2=\sigma$, then $\alpha=68.3\,\%$
      (hereafter it will be rounded to 68\%). 
\item To establish the confidence interval about $\theta$, it is said, 
      we {\it invert the probabilistic content of (\ref{eq:P_est})}:
      \begin{eqnarray}
     \mbox{from} \ \ \left\{\begin{array}{ccccc}
                     \theta & - & \Delta\theta_1 & < &  \hat\theta \\
                     \theta & + & \Delta\theta_2 & > &  \hat\theta
                    \end{array}\right.  
      && \mbox{follows} \ \  \left\{\begin{array}{ccccc}
                     \theta & < &\hat\theta  & + &  \Delta\theta_1  \\
                     \theta & > & \hat\theta & - &   \Delta\theta_2
                    \end{array}\right.\ \ 
     \end{eqnarray}      
      thus rewriting (\ref{eq:P_est}) as 
      \begin{eqnarray}
      P(\hat\theta-\Delta\theta_2 < \theta < \hat\theta+\Delta\theta_1) &=& \alpha\,.
      \label{eq:P_theta}
      \end{eqnarray}

\end{enumerate}
Then, it usually follows some 
humorous nonsense to explain what (\ref{eq:P_theta}) is not  and
what, instead, should be. Some say that {\it this formal expression 
      does not represent a probabilistic statement about $\theta$, because
      $\theta$ is not a random variable, having a well defined value, 
      although unknown}.  Instead, it is said, $\theta_1=\hat\theta-\Delta\theta_2$
      and $\theta_2=\hat\theta-\Delta\theta_1$ are random variables. 
      Then, the meaning of (\ref{eq:P_theta}) is not probability
      that $\theta$ lies inside the interval, but rather probability
      that the interval $[\theta_1,\theta_2]$
      encloses  $\theta$, in the sense (`frequentistic coverage') that 
      if {\it we repeat an infinite number of times the experiment, 
      the true value $\theta$ (that is always the same) will
      be between $\theta_1$ and $\theta_2$ (that change from time to time)
      in a fraction $\alpha$ of the cases; i.e. in each single 
      measurement the statement $\theta_1 < \theta <\theta_2$ 
      has a probability $\alpha$ of being true.} 
      (And all people of good sense wonder what is the difference 
       between the latter statement and saying that  $\alpha$ is
      the probability that $\theta$ is between $\theta_1$ and $\theta_2$.) 

Anyhow --- and more seriously ---, 
besides all this nonsense, what matters is what finally remains 
in the mind of those who learn the prescription and 
how Eq.~(\ref{eq:P_theta}) is used in scientific questions. 
If a scientist knows $f(\hat\theta\,|\,\theta)$ 
and observes  $\hat\theta$, 
then he/she feels authorized by Eq.~(\ref{eq:P_theta})
to {\it be confident}, with confidence level $\alpha$,
 that the unknown value of 
$\theta$ is in the range 
$\hat\theta-\Delta\theta_2 < \theta < \hat\theta+\Delta\theta_1$. 
And it is a matter of fact that, in practice, all users of the
prescription  consider  Eq.~(\ref{eq:P_theta}) as 
a probabilistic statement about $\alpha$, i.e. they 
feel confident that 
$\theta$ is in that interval as he/she is confident to extract a white ball 
from a box that contains a fraction $\alpha$ of white 
balls.\footnote{What happens is that the poor teacher (see footnote 1)
at the end of the day is forced to tell that  
Eq.~(\ref{eq:P_theta}) is `in practice' a probabilistic statement
about $\theta$, perhaps adding that this is not rigorously correct
but, essentially, `it can be interpreted {\it as if}'. 
However, this is not just an understandable 
imprecision of the `poor teacher', in conflict between
good sense and orthodoxy\,\cite{Maxent98}. 
For example,
we read in Ref.~\cite{JamesRoos} (the authors are influential supporters
of the use frequentistic methods in the particle physics community):
\begin{quote}
{\sl 
When the result of a measurement of a physics quantity is published as 
$R=R_0\pm\sigma_0$ without further explanation, it simply implied
that R is a Gaussian-distributed measurement with mean $R_0$
and variance $\sigma_0^2$. 
This allows to calculate various confidence
intervals of given ``probability'', i.e. the ``probability''
P that the true value of $R$ is within a given interval.
}
\end{quote} 
(The quote marks are original and nowhere in the paper is
explained why probability is in quote marks.)
}

\section{What do we really learn 
from Eqs.\,(\ref{eq:P_est})--(\ref{eq:P_theta})?}
Let us now go through the details of the previous `proof' 
and try to understand what we initially knew and what we know after 
the `probabilistic content inversion' 
provided by Eq.~(\ref{eq:P_theta}). In particular,
we need to understand what we have really
learned about the unknown 
true value $\theta$ as we went 
through the steps (\ref{eq:P_est})--(\ref{eq:P_theta}).

Given the general assumptions, 
the statement (\ref{eq:P_est}) is certainly correct.
One would argue whether $f(\hat\theta\,|\,\theta)$ is indeed 
the `right' pdf, but this
is a different story (the issue here is just logic, as we are only
interested in logical consistency of the various statements). 
Much awareness is gained about what 
is going on in steps (\ref{eq:P_est})--(\ref{eq:P_theta}),
if we rewrite Eq.~(\ref{eq:P_est}) stating 
explicitly the basic assumption as a explicit 
condition in the  probabilistic statement,
and distinguishing the name of variable from its particular numerical value,
indicating the former with a capital letter, as customary:
\begin{eqnarray}
P[ \theta-\Delta\theta_1 < \hat\Theta < \theta+\Delta\theta_2 
   \,\,|\,\, f(\hat\theta\, |\, \theta)] &=& \alpha\,,
\label{eq:P_est_cond}
\end{eqnarray}
that, for a Gaussian distribution of $\hat\Theta$ 
[indicated by the shorthand $\hat \Theta \sim {\cal N}(\theta,\sigma)$] 
and for 
$\Delta\theta_1=\Delta\theta_2=\sigma$,  becomes
\begin{eqnarray}
P[ \theta-\sigma < \hat\Theta < \theta +\sigma 
   \,\,|\,\, \hat \Theta \sim {\cal N}(\theta,\sigma)] &=& 68\%:
\label{eq:P_est_cond_norm}
\end{eqnarray}
{\it if we know} the value of $\theta$ and the standard deviation of the 
distribution we can evaluate the probability that $\hat\Theta$ 
shall occur in the interval $[\theta-\sigma,\  \theta +\sigma]$, 
where `to know' means that $\theta$ and $\sigma$ have some numeric values, 
e.g.  $\theta=5$ and $\sigma=2$. 
I find particularly enlightening to use, for a while, 
these particular values of $\mu$ and $\sigma$, rewriting 
Eq.~(\ref{eq:P_est_cond_norm}) as
\begin{eqnarray}
P[ 5-2 < \hat\Theta < 5 +2 
   \,\,|\,\, \hat \Theta \sim {\cal N}(5,2)] &=& 68\%.
\label{eq:P_est_cond_norm_5_2}
\end{eqnarray}
Obviously, knowing $\hat\Theta \sim {\cal N}(5,2)$, we can 
write an infinite number of probabilistic statements. In particular
\begin{eqnarray}
P[\hat\Theta<5-2 \,\,|\,\, \hat \Theta \sim {\cal N}(5,2)] &=& 16\%\\
P[\hat\Theta>5+2 \,\,|\,\, \hat \Theta \sim {\cal N}(5,2)] &=& 16\%\,,
\end{eqnarray}
i.e. 
\begin{eqnarray}
P[\hat\Theta+2<5 \,\,|\,\, \hat \Theta \sim {\cal N}(5,2)] &=& 16\%\\
P[\hat\Theta-2>5 \,\,|\,\, \hat \Theta \sim {\cal N}(5,2)] &=& 16\%\,,
\end{eqnarray}
from which we have\footnote{I read once in a frequentistic book
something like this:
``{\it if you do not trust logic, prove it with a Monte Carlo}''.
These are the two lines of R code~\cite{R} 
needed to `prove' by Monte Carlo the equality of 
Eqs.~(\ref{eq:P_est_cond_norm_5_2}) and 
(\ref{eq:P_est_cond_norm_inv}):\\
{\tt 
x $\leftarrow$ rnorm(10000, 5, 2)\\
length(x[ (5-2) < x \& x < (5+2) ]) == length(x[ (x-2) < 5 \& 5 < (x+2) ])\,.
} 
Similarly, for an asymmetric interval, e.g.  $\Delta\theta_1=2$
and $\Delta\theta_2=3$, we have\\
{\tt
length(x[ (5-2) < x \& x < (5+3) ]) == length(x[ (x-3) < 5 \& 5 < (x+2) ])\,.
}
}
\begin{eqnarray}
P[ \hat\Theta-2 < 5 < \hat\Theta+2
   \,\,|\,\, \hat \Theta \sim {\cal N}(5,2)] &=& 68\%.
\label{eq:P_est_cond_norm_inv}
\end{eqnarray}
Obviously, this expression is valid for any \underline{known}
value of $\theta$ and $\sigma$:
\begin{eqnarray}
P[ \hat\Theta-\sigma < \theta < \hat\Theta+\sigma
   \,\,|\,\, \hat \Theta \sim {\cal N}(\theta,\sigma)] &=& 68\%\,.
\label{eq:P_est_cond_norm_inv_sym}
\end{eqnarray}
The fact that we have replaced the numbers 5 and 2 by the generic symbols 
$\theta$ and $\sigma$ changes nothing about meaning and possible 
use of (\ref{eq:P_est_cond_norm_inv_sym}): {\it just a rephrasing of a 
probabilistic statement about $\hat\Theta$!} 

This is also the meaning of the `probabilistic content inversion' 
(\ref{eq:P_est})--(\ref{eq:P_theta}): a simple rephrasing of a probabilistic
statement about `$\hat\theta$' 
(that indeed should be $\hat\Theta$)  
under the assumption that we know the value of $\theta$
and the pdf of  `$\hat\theta$' around $\theta$:
\begin{eqnarray}
P[\theta-\Delta\theta_1 < \hat\Theta < \theta+\Delta\theta_2
 \,\,|\,\, f(\hat\theta\,|\,\theta)] &=& \alpha \label{eq:P_est_cond1} \\
\mbox{equivalent to}\hspace{6cm} && \hspace{0.5cm}\mbox{} \nonumber \\
P[\hat\Theta-\Delta\theta_2 < \theta < \hat\Theta+\Delta\theta_1
 \,\,|\,\, f(\hat\theta\,|\,\theta)] &=& \alpha  \label{eq:P_theta_cond}
\end{eqnarray}
Therefore, there is no doubt that (\ref{eq:P_theta_cond}) follows from 
(\ref{eq:P_est_cond1}). The question is that this is true if 
$\theta$, $\Delta\theta_1$ and  $\Delta\theta_2$ are real numbers,
whose values, together with the knowledge of $f(\hat\theta\,|\,\theta)$,
allow us to calculate $\alpha$. 
However, rephrasing the probabilistic statement concerning 
the possible observation $\hat\Theta$ given a certain $\theta$ 
does not help us in solving the problem we are really interested in, 
i.e. to gain knowledge about the true value and to express our  
level of confidence on it, given 
the experimental observation $\hat\theta$.

What we are really looking for is, indeed,
$P(\theta_1 < \Theta < \theta_2   \,\,|\,\, \hat\theta)$.
But this can be only achieved if we are able to write down, 
given the best knowledge of the physics case, the pdf 
$f(\theta\,|\,\hat\theta)$.
Pretending to express our confidence 
about $\Theta$ without passing through $f(\theta\,|\,\hat\theta)$
is pure nonsense, based on a proof that reminds the 
`game of the three cards' 
proposed by con artists in disreputable streets.
Now, it is a matter of logic that 
the only way to go from  $f(\hat\theta\,|\,\theta)$ to 
$f(\theta\,|\,\hat\theta)$ is to make a
correct `probability inversion',
following the rules of probability theory,
instead of that shameful outrage against logic.
 The probabilistic
tool to perform the task is Bayes' theorem, by which
it is possible to establish intervals that contain the true
value at a given level of {\it probability} (meant really
by how much we are confident the true value is in a given interval!).
It is easy to show that, under well defined conditions that
often hold in routine applications, the interval
calculated at a given level of probability $\alpha_p$ 
is equal to `confidence interval' calculated with a `confidence
level' $\alpha_{CL}$, if numerically\footnote{Those interested in 
Bayesian/frequentistic comparisons
might give a look at Ref.~\cite{Zech}.
Personally, 
as explained in Ref.~\cite{BR} (footnote 18 of p. 229), I dislike the 
quantitative comparisons of Bayesian and frequentistic methods to solve
the same problem simply because
quantitative comparisons assume 
that we are dealing with homogeneous quantities, 
while frequentistic CL's
and Bayesian probability intervals are as homogeneous 
as apples and tomatoes are. (This is also the reason 
I used here two different symbols, $\alpha_{CL}$ and $\alpha_p$.)}  
$\alpha_{CL}=\alpha_p$
(see e.g. Ref.~\cite{BR}).
It is not a surprise, then, that the confidence interval
prescriptions yield quite often a correct result,
but they might also miserably fail, especially 
in frontier physics applications.

\section{Conclusions}
The proof of frequentistic `confidence intervals' is sterile
and there is no logical reason why one should attach 
a `level of confidence` to the intervals calculated following that prescription.
Paraphrasing a sentence of the same author 
of the opening quote, 
{\it they are called confidence intervals by 
their advocates because they provide confidence in no other way}.

\end{document}